\documentclass[prl,aps,amssymb,twocolumn,showpacs,floatfix]{revtex4}
\usepackage[dvips]{graphicx}
\def\be{\begin{equation}}
\def\ee{\end{equation}}
\def\bea{\begin{eqnarray}}
\def\eea{\end{eqnarray}}
 
\begin{document}
 
\title{Relaxation processes in a disordered Luttinger liquid}
\author{D. A. Bagrets$^{1}$}
\author{I. V. Gornyi$^{1,2}$}
\author{A. D. Mirlin$^{1,3,4}$}
\author{D. G. Polyakov$^{1}$} 
\affiliation{
$^{1}$Institut f\"ur Nanotechnologie, Forschungszentrum Karlsruhe, 
76021 Karlsruhe, Germany 
\\
$^{2}$A.F.Ioffe Physico-Technical Institute, 
194021 St.~Petersburg, Russia
\\
$^{3}$Institut f\"ur Theorie der kondensierten Materie, Universit\"at
Karlsruhe, 76128 Karlsruhe, Germany
\\
$^{4}$Petersburg Nuclear Physics Institute, 
188300 St.~Petersburg, Russia
}

%\date{\today}

\begin{abstract} 
The Luttinger liquid model, which describes interacting electrons in a
single-channel quantum wire, is completely integrable in the absence of
disorder and as such does not exhibit any relaxation to equilibrium. We
consider relaxation processes induced by inelastic electron-electron
interactions in a disordered Luttinger liquid, focusing on the equilibration
rate and its essential differences from the electron-electron scattering rate
as well as the rate of phase relaxation. In the first part of the paper, we
review the basic concepts in the disordered Luttinger liquid at
equilibrium. These include the elastic renormalization, dephasing, and
interference-induced localization. In the second part, we formulate a
conceptually important framework for systematically studying the
nonequilibrium properties of the strongly correlated (non-Fermi) Luttinger
liquid. We derive a coupled set of kinetic equations for the fermionic and
bosonic distribution functions that describe the evolution of the
nonequilibrium Luttinger liquid. Remarkably, the energy equilibration rate in
the conducting disordered quantum wire (at sufficiently high temperature, when
the localization effects are suppressed by dephasing) is shown to be of the
order of the rate of elastic scattering off disorder, independent of the
interaction constant and temperature.
\end{abstract}
 

\maketitle

\section{I. Introduction}
\label{Intro}

Strongly correlated electron systems in one dimension (1d) have become an area
of immense interest from the perspective of both fundamental and technological
aspects of nanophysics. Intense experimental effort has focused on such
realizations of quantum wires with a few or single conducting channels as
cleaved-edge \cite{auslaender02}, V-groove \cite{palevski05}, and
crystallized-in-a-matrix \cite{zaitsev-zotov00} semiconductor quantum wires,
coupled quantum Hall edges running in opposite directions
\cite{kang00,grayson05}, single-wall carbon nanotubes \cite{nt}, polymer
nanofibers \cite{aleshin04}, and metallic nanowires
\cite{slot04,venkataraman06}. Central to much of the fascinating physics of
the 1d systems is that electron-electron (e-e) interactions in 1d geometry can
have dramatic effects leading to the emergence of a Luttinger liquid (LL)
\cite{giamarchi04}. The latter constitutes a canonical example of a non-Fermi
liquid, in which quasiparticle fermionic excitations are inappropriate to
describe low-energy physics. At the foundation of the conventional LL theory
is a description in terms of bosonic elementary excitations (plasmons,
spinons) \cite{giamarchi04}. Following this approach, the ground-state
properties of a clean LL are well understood for arbitrary strength of
interaction. Much has also been learned about the LL in the presence of a
single compact scatterer \cite{kane92,giamarchi04}. However, as far as a
disordered LL is concerned, a number of important questions, even at the most
fundamental level, remained largely unanswered until very recently (for a
review see Ref.~\cite{gornyi07}).

In the presence of disorder, quantum interference of scattered electron waves
leads to the effects of Anderson localization \cite{anderson58}. Similarly to
e-e interactions, the lower the dimensionality, the stronger the localization
effects. In a 1d electron gas, all electron states are localized even for an
arbitrarily weak random potential and the localization length is the mean free
path. In the case of noninteracting electrons, the quantum localization in 1d
has been studied in great detail (see, e.g., Ref.~\cite{berezinskii73}). A
principal complication that arises in the disordered LL is that the quantum
interference phenomena yielding the Anderson localization are conventionally
treated in terms of fermions, by employing the concepts of interference and
dephasing of fermionic excitations. The question of to what extent the notion
of phase relaxation in the localization problem is applicable to the
(non-Fermi) LL is therefore of crucial importance. Recently, this problem was
addressed in Refs.~\cite{gornyi05,gornyi07}, where the interaction-induced
dephasing rate that governs the localization term in the conductivity of the
disordered LL was calculated.

Another conceptually nontrivial aspect of the interplay between disorder and
interaction concerns the nonequilibrium properties of the LL. In the
homogeneous case, the LL model is completely integrable and as such does not
exhibit any relaxation to equilibrium: an arbitrary excited state will never
decay to the equilibrium state characterized by temperature. Of central
importance is therefore the question of how the equilibration of fermionic
and/or bosonic excitations in the LL occurs in the presence of disorder.

This paper is primarily concerned with various relaxation processes associated
with inelastic interactions between electrons in the disordered LL.
Specifically, we focus on the rates of e-e scattering, phase relaxation, and
energy relaxation, with emphasis on the essential differences between them. In
%Sec.~\ref{mod}
Sec.~II
we begin with the formulation of the model. Section 
%\ref{dru}
III
highlights a few aspects of temperature-dependent screening of disorder in
1d. Section 
\ref{deph} 
IV
covers the problem of phase relaxation---this
discussion largely follows the results of
Refs.~\cite{gornyi05,gornyi07} and serves as the starting point for our
approach to nonequilibrium physics of the LL. In 
%Sec.~\ref{erel} 
Sec.~V
we consider
energy relaxation and introduce a general framework \cite{bagopo} for studying 
the behavior of the disordered LL out of equilibrium.

\section{II. Model}
\label{mod}

Let us specify the model. By decomposing the electron operator into right- and
left-moving parts , $\psi(x)=\psi_+(x)+\psi_-(x)$, the Hamiltonian of a
disordered LL is written as
\bea
H&=&H_{\rm kin}+H_{\rm ee}+H_{\rm dis}~,
\label{1}\\
H_{\rm kin}&=&
-v_F\sum_{\mu=\pm}\int \!dx\,
\psi^\dagger_{\mu}\left(i\mu\partial_x+k_F\right) \psi_{\mu}~,
\label{2}
\\
H_{\rm ee}&=&{1\over 2}\sum_{\mu=\pm}\int \!dx
\left( n_{\mu} \, V_f\,
n_{-\mu}
+ n_{\mu} \, \tilde{V}_f\, n_{\mu}
\right)~,
\label{3} 
\\
H_{\rm dis}&=& \int \!dx
\left[ U_b(x)\ \psi^\dagger_-\psi_++{\rm H.c.}\right]~.
\label{4}
\eea
Here $n_\mu= \psi^\dagger_\mu \psi_\mu$ is the density of the right and left
movers and their dispersion relation is linearized about two Fermi points at
the wavevectors $\pm k_F$ with the velocity $v_F$. Throughout the paper we
consider spinless electrons (for spin-related effects see
Ref.~\cite{yashenkin}).

The e-e interaction, Eq.~(\ref{3}), is characterized by the Fourier components
of the short-range (screened) interaction potential with zero momentum
transfer $V_f$ (forward scattering between right and left movers) and
$\tilde{V}_f$ (forward scattering of electrons from the same chiral branch on
each other). Unless the right and left movers are spatially separated (as in
coupled quantum Hall edges), $\tilde{V}_f=V_f$. The Luttinger model {\it per
se} does not include backward scattering characterized by the Fourier
component $V_b$ with momentum transfer $\pm 2k_F$. For spinless electrons,
however, $V_b$ can be trivially incorporated by shifting $V_f\to V_f-V_b$,
since two types of scattering---due to $V_f$ and $V_b$---are then related to
each other as direct and exchange processes. The local interaction between
identical fermions $\tilde{V}_f$ yields no scattering, but, due to a quantum
anomaly in the LL model, generates a shift of the Fermi velocity $v_F\to
v_F^*=v_F+\tilde{V}_f/2\pi$. It is customary to parametrize the strength of
e-e interaction by means of the Luttinger constant $K$:
\be
K=\left({1-\alpha\over 1+\alpha}\right)^{1/2}~,\quad \alpha={V_f\over 2\pi
v_F^*}~.
\label{5}
\ee
The velocity of elementary excitations (plasmons) in a clean LL is given by
\be
u=v_F^*(1-\alpha^2)^{1/2}~,
\label{5a}
\ee
which transforms for $\tilde{V}_f=V_f$ into $u=v_F/K$. 

The low-energy theory described by the Hamiltonian (\ref{1}) is only then
well-defined when supplemented by an ultraviolet energy cutoff $\Lambda$. The
latter depends on microscopic details of the problem and obeys
\be
\Lambda=u/\pi\lambda~,
\ee
where the length scale $\lambda$ is set by the lattice constant, the Fermi
wavelength, or the spatial range of interaction in the original microscopic
theory, whichever gives the smallest $\Lambda$. Thus the complete set of
parameters defining the LL model in the absence of disorder includes $v_F^*$,
$V_f$, and $\Lambda$. It is worth noting that the input parameters of the
low-energy theory include Fermi-liquid-type renormalizations coming from
energy scales larger than $\Lambda$; in particular, the ``bare" $v_F$ in
Eq.~(\ref{1}) in general is not an interaction-independent constant
if the interaction is strong ($1-K\sim 1$).

The term $H_{\rm dis}$, Eq.~(\ref{4}), describes backscattering of electrons
off a static random potential $U(x)$. The latter is taken to be of white-noise
type with the correlators of backscattering amplitudes
\be
\overline{U_b(x)U_b^*(0)}=\overline{U(x)U(0)}=w\delta(x)
\label{6}
\ee
and $\overline{U_b(x)U_b(0)}=0$. The forward-scattering amplitudes are omitted
in Eq.~(\ref{4}) since they can be gauged out in the calculation of the
conductivity.  

\section{III. Elastic scattering}
\label{dru}

One of the characteristic features of a LL is a large renormalization of the
strength of disorder (\ref{6}) by e-e interaction. In particular, the
conductivity without any localization \cite{anderson58} or pinning
\cite{larkin70} effects included (``Drude
conductivity") is $\sigma_{\rm D}(\omega,T)=e^2v_F/\pi[-i\omega+M(\omega,T)]$,
where the disorder-induced scattering rate in the dc limit
\be
{1\over \tau(T)}={\rm Re}\,M(0,T)
=a_K\,{1\over \tau_0}\left({\Lambda\over T}\right)^{2(1-K)}
\label{7}
\ee
grows as a power law with decreasing temperature $T$ for repulsive interaction
($K<1$). The momentum relaxation rate in the absence of interaction is given
by $\tau_0^{-1}=2wv_F^{-1}$ with $w$ from Eq.~(\ref{6}). Calculating the Drude
conductivity at finite $\omega$ and sending $\omega\to 0$ afterwards allows to
unambiguously determine the coefficient \cite{mirlin07}
$a_K=\Gamma^2(1+K)/\Gamma (2K)$ in the relaxation rate. Here and below the
disorder is supposed to be weak in the sense that $\Lambda\tau\gg 1$.

The underlying physics of the renormalization (\ref{7}) can be described in
terms of the $T$-dependent screening of individual impurities; specifically,
in terms of scattering by Friedel oscillations which slowly decay in real
space and are cut off on the spatial scale of the thermal length. At this
level, the only peculiarity of the LL as compared to higher dimensionalities
is that the renormalization of $\tau$ is more singular and, most importantly
from the calculational point of view, necessitates going beyond the
Hartree-Fock approximation, even for weak interaction (see, e.g.,
Ref.~\cite{polyakov03}).

In general, not only the strength of disorder but also the strength of
interaction is subject to renormalization and depends on $T$, so that the
function $\tau(T)$ is not a simple power law. An important question,
therefore, is under what condition the exponent in Eq.~(\ref{7}) is given by
the bare interaction coupling constant. One of the approaches to the problem
was formulated in Ref.~\cite{giamarchi88} in terms of a bosonic
renormalization group (RG). The RG approach does not allow to obtain the
$K$-dependent prefactor $a_K$ in Eq.~(\ref{7}), but is particularly beneficial
in predicting the $T$ dependence of the Drude conductivity. For spinless
electrons, the one-loop RG equations read
\bea
dK/d{\cal L}&=&f(K){\cal D}~,
\label{8a}
\\
d{\cal D}/ d{\cal L}&=&(3-2K){\cal D}~,
\label{8b}
\eea
where ${\cal L}=\ln L/\lambda$ and ${\cal D}=2w\lambda/\pi u^2$. For the Drude
conductivity (i.e., as long as the localization effects are not included, see
Sec.~IV), the spatial scale $L$ is given by the thermal length $u/T$. The
scattering rate $1/\tau (T)$ is then proportional to $T{\cal D}(T)$. The
function $f(K)=-K^2/2+(1+K^2)(3-2K)/4$ vanishes at $K=1$, so that interaction
is not generated by disorder (in the original equations of
Ref.~\cite{giamarchi88}, the coupling constant $K$ contains an admixture
of disorder and therefore the corresponding $f(1)\neq 0$, see
Ref.~\cite{gornyi07} for a discussion of this point); moreover, the
interaction (hence $1-K$) does not change sign in the course of
renormalization.

The RG flow (\ref{8a}),(\ref{8b}) is characterized by a separatrix which
behaves as ${\cal D}=8(K-3/2)^2/9$ for $K>3/2$ and terminates at $K=3/2$. For
the bare (taken at $L=\lambda$) values of $\cal D$ and $K$ that lie below the
separatrix (i.e., for the case of strong attractive interaction with $K>3/2$),
the disorder strength $\cal D$ renormalizes to zero, otherwise $\cal D$ grows
with increasing $L$ to a strong-coupling point with ${\cal D}\sim 1$. The
renormalization of the coupling constant $K$ by disorder is essential if the
RG trajectory is close to the parabola ${\cal D}=8(K-3/2)^2/9$. For example,
if the RG flow passes through the point $K=3/2$, the integration of
Eqs.~(\ref{8a}) and (\ref{8b}) gives for ${\cal D}\ll 1$:
\be
{\cal D}-{\cal D}_0={\cal D}_0\,\tan^2\left({3{\cal D}_0^{1/2}\over
2^{3/2}}\ln {L\over L_0}\right)~,
\label{9}
\ee 
where ${\cal D}_0$ and $L_0$ are the values of ${\cal D}$ and $L$ at $K=3/2$
and the sign of $\ln (L/L_0)$ is positive for running $K<3/2$ and negative
otherwise. One sees that ${\cal D}$ grows with increasing $L$ for $K<3/2$ as
\be
{\cal D}=8/9\ln^2(l/L)
\label{10}
\ee
(for ${\cal D}_0\ll {\cal D}\ll 1$). Here the renormalized mean free path $l$
(the scale at which ${\cal D}\sim 1$) obeys $\ln (l/L_0)=2^{1/2}\pi/3{\cal
D}_0^{1/2}$. The logarithmic dependence of $\cal D$ on $L$ is precisely due to
the renormalization of $K$.

On the other hand, if the bare $K<3/2$, the RG trajectory follows
Eq.~(\ref{9}) with $L_0=\lambda$ and ${\cal D}_0$ understood as the bare value
of ${\cal D}$ only at ${\cal D}-{\cal D}_0\gg (K-3/2)^2$. Integrating
Eqs.~(\ref{8a}),(\ref{8b}) in the opposite limit 
\be
{\cal D}-{\cal D}_0\ll
(K-3/2)^2~,
\label{11}
\ee
one gets 
\be
{\cal D}={\cal D}_0(L/\lambda)^{3-2K}~,
\label{12}
\ee
which corresponds to Eq.~(\ref{7}). Equation (\ref{11}) thus answers the
question of when the renormalization of $K$ may be neglected. Notice that for
repulsive interaction ($K<1$) the condition (\ref{11}) is satisfied for the
whole range of ${\cal D}\ll 1$ [which is where the RG equations
(\ref{8a}),(\ref{8b}) are valid]. It follows that for the most relevant case
of direct Coulomb interaction the renormalization of interaction on ballistic
scales (${\cal D}\ll 1$) plays no role and the exponent in Eq.~(\ref{7}) is
$T$-independent and given by the bare value of $K$ (the one in a clean
system). In other words, the renormalization of disorder for repulsive
interaction reduces to the renormalization of an individual impurity. It is
worth emphasizing that this does not mean that the disorder-induced correction
to the bare value of $1-K$ is small: in fact, the correction is of the order
of $1-K$ itself when ${\cal D}\sim 1$. The point is that the exponent of
${\cal D}(L)$ and, correspondingly, of the renormalized scattering rate
$1/\tau(T)$ is not given by the running coupling constant $K$, but rather is
accumulated on the whole RG trajectory.

\section{IV. Phase relaxation}
\label{deph}

The renormalization of $\tau$ stops with decreasing $T$ at 
\be 
T\tau(T)\sim 1~,
\label{13} 
\ee 
since the long-range Friedel oscillations created by disorder are cut off even
at zero $T$ on the spatial scale of the disorder-induced mean free path. This
condition gives the zero-$T$ mean free path $l\propto \tau_0^{1/(3-2K)}$
[notice that Eq.~(\ref{13}) is also expressible as ${\cal D}(L)\sim 1$ with
$L=u/T$] and, correspondingly, the zero-$T$ localization length $\xi\sim
l$. It is important to stress, however, that the above condition does not
correctly predict the onset of localization with decreasing $T$---in contrast
to the argument, frequently stated in the literature (see, e.g.,
Ref.~\cite{giamarchi04} and references therein) and based on the RG equations
(\ref{10}),(\ref{11}), which treat scalings with the length scales $L$ and
$u/T$ as interchangeable. While substituting $u/T$ for $L$ is justified for
the ``elastic renormalization" [Eq.~(\ref{7})], the one-loop equations
(\ref{10}),(\ref{11}) miss, by construction, the interference effects
(coherent scattering on several impurities) that lead to localization. The
status of the RG \cite{giamarchi88} is thus that of the Drude formula for
interacting electrons. The $T$ dependence of the conductivity $\sigma(T)$,
however, comes not only from the $T$-dependent screening of disorder
[Eq.~(\ref{7})], but also from the localization term in $\sigma(T)$ whose
amplitude is governed by phase relaxation due to inelastic e-e scattering. The
temperature below which the localization effects become strong is, in contrast
to Eq.~(\ref{13}), determined by the condition
\be
\tau(T)/\tau_\phi(T)\sim 1~,
\label{14}
\ee
where $\tau_\phi$ is the weak-localization dephasing time. Notice that for
weak interaction ($1-K\simeq \alpha\ll 1$), Eq.~(\ref{14}) is satisfied at
much higher $T$ than Eq.~(\ref{13}). Below we introduce the notion of
dephasing of localization effects in the disordered LL and analyze the phase
relaxation in the limit of weak interaction.

The very applicability of the notion of dephasing, as we know it from the
studies of higher-dimensional Fermi-liquid systems, to the LL is not
altogether apparent. A subtle question concerns the nature of elementary
excitations in the LL, especially in the presence of disorder. The clean LL is
a completely integrable model which is represented in terms of noninteracting
(hence nondecaying) bosons; however, the phase relaxation in electron systems
is conventionally described in terms of interacting fermions. Physically, the
difficulty is related to the fact that the bosonized approach describes
propagation of density fluctuations, whereas the natural language for quantum
interference phenomena is that of quantum amplitudes. To study the
interference effects and their dephasing, one has therefore to either proceed
with the standard bosonization, poorly suited to describe the quantum
interference in the inhomogeneous case, or try to define the observables in
such a way that they can be expressible in terms of decaying fermionic
excitations. In what follows in this section, we take the latter path and give
a succinct analysis of the phase relaxation in the disordered LL, based on the
results obtained within the ``functional bosonization" formalism
\cite{gornyi07} and the quasiclassical formalism \cite{gornyi05}, both of
which combine the fermionic and bosonic approaches to the problem.

Let us first point out one of the subtleties of the LL model, which is crucial
to our discussion of the phase and energy relaxation. The Golden rule
expression for the e-e collision rate at equilibrium, as follows from the
Boltzmann kinetic equation, reads
\bea
{1\over \tau_{\rm ee} (\epsilon)}&=&\int\! d\omega\int\!  d\epsilon' \,
{\cal K}(\omega)\nonumber \\
&\times&
 \left(f^h_{\epsilon-\omega}f_{\epsilon'}
f^h_{\epsilon'+\omega}+f_{\epsilon-\omega}
f^h_{\epsilon'}f_{\epsilon'+\omega}\right)~,
\label{15}
\eea
where $f_\epsilon$ is the Fermi distribution function and
$f^h_{\epsilon}=1-f_{\epsilon}$. Consider the {\it clean} case. Then the
scattering kernel ${\cal K}(\omega)={\cal K}^H_{++}(\omega)+{\cal
K}^H_{+-}(\omega)+{\cal K}^F(\omega)$ to second order in the interaction is
given by
\bea
{\cal K}^H_{++}&=&{\tilde{V}_f^2\over \pi^3\rho}\int \!{dq\over 2\pi}\, 
\left[\,{\rm Re} D_{+}(\omega,q)\,\right]^2~,
\label{16} 
\\
{\cal K}^H_{+-}&=&
{V_f^2\over \pi^3\rho}\int \!{dq\over 2\pi} \,
 {\rm Re} D_{+}(\omega,q)\,{\rm Re} D_{-}(\omega,q)~,
\label{17} 
\eea
and ${\cal K}^F=-{\cal K}^H_{++}$. The Hartree terms ${\cal K}^H_{++}$ and
${\cal K}^H_{+-}$ are related to scattering of two electrons from the same or
different chiral spectral branches, respectively, ${\cal K}^F$ is the Fock
(exchange) counterpart of ${\cal K}^H_{++}$, the thermodynamic density of
states $\rho=1/\pi v_F$, and $D_\pm=i\pi \rho/(\omega\mp qv_F+i0)$ are the
two-particle propagators in the clean limit.

Substituting Eqs.~(\ref{16}),(\ref{17}) in Eq.~(\ref{15}), we obtain the
lowest-order result for the e-e scattering rate at the Fermi level
$(\epsilon=0)$ in terms of the corresponding contributions to the retarded
electronic self-energy $\Sigma_+$ defined by
$G^R_+(\epsilon,p)=[\,\epsilon-v_Fp-\Sigma_+(\epsilon,p)\,]^{-1}$, where
$G^R_+$ is the retarded Green's function for right-movers. Specifically,
$\tau^{-1}_{\rm ee}=-2{\rm Im}\,\Sigma_+(0,0)$ with
$\Sigma_+(0,0)=\Sigma^H_{++}+\Sigma^H_{+-}+\Sigma^F$, where 
\bea 
{\rm Im}\Sigma^H_{+\pm}&=&-{\pi\over 2}\alpha^2v_F\!\int \!d\omega
\,\omega\left(\coth{\omega\over 2T}-\tanh{\omega\over 2T}\right)\nonumber \\
&\times&\int \! dq\, \delta(\omega-v_Fq)\delta(\omega\mp v_Fq)~, 
\label{18}
\eea 
$\Sigma^F=-\Sigma^H_{++}$, and we put $V_f=\tilde{V}_f$. One sees that the
contribution of $\Sigma^H_{++}$ is diverging. For spinless electrons, however,
the divergency is canceled by the exchange interaction. Indeed, as we have
discussed in Sec.~II, the $\tilde{V}_f$ interaction drops out of the
problem in this case, inducing only a shift of the velocity $v_F\to v_F^*$. It
is worth noting that the ``Hartree-Fock cancellation" is only exact for the
point-like interaction (when $\tilde{V}_f$ is independent of the transferred
momentum), otherwise $\tilde{V}_f$ yields a nonzero contribution
\cite{chalker07} to $\tau_{ee}^{-1}$. The latter is small and can be neglected
in the low-$T$ limit for $\tilde{V}_f=V_f$ but is the only one present for
$V_f=0$, which is the case, e.g., for an isolated quantum Hall edge. The
remaining term $\Sigma^H_{+-}$ gives
\be
\tau_{ee}^{-1}=-2{\rm Im}\,\Sigma^H_{+-}=\pi\alpha^2T~.
\label{19}
\ee

This may look very similar to the familiar $T^2$ or $T^2\ln T$ dependence of
the e-e scattering rate in clean three- or two-dimensional electron systems,
respectively. However, the nontrivial point---which demonstrates the
peculiarity of the LL model---is that $\tau_{ee}^{-1}$ in Eq.~(\ref{19}) is
determined by 
\be 
\omega, q=0~, 
\label{20} 
\ee
i.e., by scattering processes with infinitesimally small energy transfers, in
contrast to higher dimensions where the characteristic transfer is of order
$T$. On the other hand, it is worth emphasizing that $T\tau_{ee}\gg 1$ for
$\alpha\ll 1$, which in Fermi-liquid theory is commonly referred to as one of
the conditions for the existence of a Fermi liquid. In this respect, the
weakly interacting LL, while being a canonical example of a non-Fermi liquid,
reveals the typical Fermi-liquid property. The LL physics (e.g., the power-law
singularity of the tunneling density of states at the Fermi level) is in fact
encoded in the singular {\it real} part of the self-energy $\Sigma_+(\epsilon,
p)$ (for more details see Ref.~\cite{gornyi07}).

It is the property (\ref{20}) that actually makes the 1d case special as far
as the dephasing problem is concerned. Indeed, in the spirit of
Ref.~\cite{altshuler82}, soft inelastic scattering with $qv_F,\omega\ll
\tau_\phi^{-1}$ is not expected to contribute to the dephasing of the
localization effects. In higher dimensions, in the ballistic regime $T\tau\gg
1$, this infrared cutoff is of no importance and the dephasing rate
$\tau_\phi^{-1}$ is given \cite{narozhny02} by $\tau_{ee}^{-1}$. However, in
view of Eq.~(\ref{20}), $\tau_\phi^{-1}$ in 1d cannot possibly reduce to
$\tau_{ee}^{-1}$.

\begin{figure}[ht] 
\centerline{
\includegraphics[width=6cm]{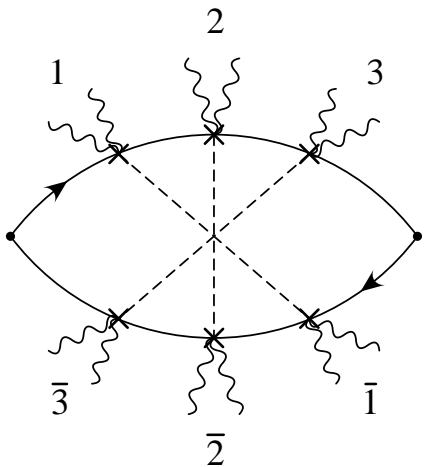}
}
\caption{
Three-impurity Cooperon diagram with interaction effects encoded in the
fluctuating factors $\exp (\pm\theta_\mu)$ (denoted by the wavy lines)
attached to the backscattering vertices (marked by the crosses). The dashed
lines connect the backscattering vertices belonging to the same impurity
(e.g., 1 and $\bar{1}$ refer to two backscatterings off impurity 1 at two 
different times).
}\label{f1} 
\end{figure}

The dephasing rate $\tau_\phi^{-1}$ can be accurately defined by calculating
the weak-localization correction to the conductivity of the disordered LL as a
function of $T$ \cite{gornyi05,gornyi07}. The leading localization correction
$\Delta\sigma$ in the ballistic limit $\tau_\phi/\tau\ll 1$ is related to the
interference of electrons scattered by three impurities. One of the diagrams
contributing to $\Delta\sigma$ (for the complete set of diagrams at the
leading order see Ref.~\cite{gornyi05,gornyi07}) is given by a
``three-impurity Cooperon" (Fig.~\ref{f1}), which describes the propagation of
two electron waves along the path connecting three impurities (``minimal
loop") in time-reversed directions. In the absence of dephasing, quantum
interference processes involving a larger number of impurities sum up to
exactly cancel (similarly to the noninteracting case \cite{berezinskii73}) the
Drude conductivity $\sigma_{\rm D}=e^2v_F\tau/\pi$, where $\tau$ is given by
Eq.~(\ref{7}). For $\tau_\phi/\tau\ll 1$, they only yield subleading
corrections through a systematic expansion in powers of $\tau_\phi/\tau$.

Within the functional-bosonization description of the LL
\cite{fogedby76,lerner04}, extended in Ref.~\cite{gornyi07} to treat
disordered systems, the interaction can be exactly accounted for by performing
a local gauge transformation $\psi_\mu(x,\tau)\to \psi_\mu(x,\tau)\exp
[\,i\theta_\mu(x,\tau)\,]$, where the bosonic field $\theta_\mu(x,\tau)$ is
related to the Hubbard-Stratonovich decoupling field $\varphi(x,\tau)$ by
\be
(\partial_\tau-i\mu
v_F\partial_x)\theta_\mu(x,\tau)=\varphi (x,\tau)~.
\label{20a}
\ee
Here $\tau$ is the Matsubara time. The correlator
$\left<\varphi(x,\tau)\varphi(0,0)\right>=V(x,\tau)$ gives the dynamically
screened interaction, for which the random-phase approximation (RPA) in the LL
model is exact \cite{dzyaloshinskii74}. In the presence of impurities, the
interaction can thus be completely gauged out to the backscattering impurity
vertices---Eq.~(\ref{20a}) is then {\it exact} for any given realization of
the impurity potential. In Fig.~\ref{f1}, the fluctuating disorder-induced
gauge factors are denoted by the wavy lines attached to the backscattering
vertices: each impurity vertex contributes the factor $\exp [\pm
(\theta_+-\theta_-)]$ and the averaging over fluctuations of $\varphi$ pairs
all the fields $\theta_\pm$ with each other.  The interaction thus induces the
factor
\be
\exp (-S_C)=\left<\exp [\,i(\theta_f-\theta_b)\,]\right> 
\label{21a}
\ee
to the Cooperon loop, where $\theta_f$ and $\theta_b$ are the phases
accumulated by an electron propagating along the ``forward'' and ``backward''
paths and the averaging is performed over the fluctuations of the field
$\varphi$. Notice that the averaging couples with each other not only the
phases $\theta_\pm$ attached to the impurities shown in Fig.~\ref{f1} but also
those attached to impurities which yield damping of the dynamically screened
interaction. As shown in Refs.~\cite{gornyi05,gornyi07}, the boson
damping is crucially important for the dephasing (see below) and a
parametrically accurate approximation is to include impurity-induced
backscattering in the effective interaction at the level of the
disorder-dressed RPA (``dirty RPA"). The total Cooperon action
\be
S_C=S+S_{\rm renorm}
\label{21b}
\ee
accounts then for both the dephasing ($S$) and the elastic renormalization of
impurities ($S_{\rm renorm}$) and we refer the reader for technical details of
the formalism to Ref.~\cite{gornyi07}.  

The leading localization correction to the conductivity can be represented in
the form \cite{gornyi05,gornyi07}
\be
\Delta\sigma=-2\sigma_{\rm D}\!\int_0^\infty \!\!dt_c \!\int_0^\infty
\!\!dt_a P_2(t_c,t_a)\exp \left[-S(t_c,t_a)\right]~, 
\label{21} 
\ee 
where 
$P_2(t_c,t_a)=(1/8\tau^2)\exp (-t_c/2\tau)\Theta(t_c-2t_a)$ is the probability
density of return to point $x=0$ after two reflections at points $x=ut_a$ and
$x=-u(t_c/2-t_a)$. Here $ut_c$ gives the total length of the Cooperon loop and
$ut_a$, being the distance between two rightmost impurities, parametrizes the
geometry of the loop. The classical trajectory for the Cooperon is
characterized by a {\it single} velocity \cite{gornyi07} and this is $u$ (the
difference between $u$ and $v_F$ can be ignored for $\alpha\ll 1$, but
uniformly on the whole trajectory). The phase relaxation is encoded in the
dephasing action $S$ in Eq.~(\ref{21}), which is a growing function of the
size of the Cooperon loop and cuts off the localization correction at $t_c\sim
\tau_\phi$. The dephasing rate $\tau_\phi^{-1}$ is thus defined by the
characteristic scale of $t_c$ on which the dephasing action $S\sim 1$.

In the limit $t_c\ll\tau$ the action reads \cite{gornyi05,gornyi07}:
\be  
S(t_c,t_a)=2\pi\alpha^2 \,T\,t_a \left(t_c-2t_a\right)/\tau~.
\label{22}
\ee
Substitution of Eq.~(\ref{22}) into Eq.~(\ref{21}) gives
\be
\Delta\sigma = -{1\over 4}\,\sigma_{\rm D}\left(\tau_\phi\over
\tau\right)^2\!\ln{\tau\over \tau_\phi} \propto {1\over
\alpha^2T}\,\ln(\alpha^2T)~,
\label{23}
\ee
where
\be
\tau_\phi^{-1}=\alpha(\pi T/\tau)^{1/2}~.
\label{24}
\ee
One sees that the phase relaxation in the disordered LL occurs on
time scales much longer than the lifetime $\tau_{ee}$: 
\be
\tau_\phi=(\tau_{ee}\tau)^{1/2}\gg \tau_{ee}~.
\label{25}
\ee
Note that $\tau_\phi^{-1}$ vanishes in the clean limit, in contrast to the
total e-e scattering rate---in agreement with the observation (\ref{20}) and
the basic fact \cite{altshuler82} that scattering with energy transfers
smaller than $\tau_\phi^{-1}$ is not effective in dephasing the localization
effects.

The vanishing of the dephasing action at $\tau^{-1}\to 0$ can be made more
transparent from the technical point of view by representing $S$ as a
difference between the self-energy ($S_{\rm ff}+S_{\rm bb}$) and vertex
($S_{\rm fb}+S_{\rm bf}$) contributions. Here the terms $S_{ij}$, associated
with an inelastic interaction between electrons propagating along the paths
$x_i(t)$ and $x_j(t)$, where $i,j=({\rm f})$ and $({\rm b})$ stand for the
``forward" and ``backward" time-reversed paths in the Cooperon, are given by
\bea
S_{ij}&=&-T\int
{d\omega\over 2\pi}\int {dq\over 2\pi}\int_0^{t_c} \!dt_1\!\int_0^{t_c} \!dt_2
\,{1\over \omega}\,{\rm Im}\,V_{\mu\nu}(\omega,q) \nonumber \\ &\times&\,\,\exp
\{iq\left[\,x_i(t_1)-x_j(t_2)\,\right]-i\omega(t_1-t_2)\}~. 
\label{26}
\eea{eqnarray} 

Equation (\ref{26}) is similar to that in higher dimensionalities (``AAK
action" \cite{altshuler82}) with one subtle and important distinction.
Because of the Hartree-Fock cancellation of the bare interaction $\tilde{V}_f$
between electrons from the same chiral branch [recall the discussion after
Eq.~(\ref{18})], the dynamically screened retarded interaction
$V_{\mu\nu}(\omega,q)$ acquires the indices $\mu,\nu$ denoting the direction
of motion of the interacting electrons: $\mu={\rm sgn}\,\dot{x}_i$, $\nu={\rm
sgn}\,\dot{x}_j$. If one would keep both $V_f$ and $\tilde{V}_f$ processes in
$V(\omega,q)$, the dephasing action in 1d could {\it not} be written in the
form of Eq.~(\ref{26})---in contrast to higher dimensionalities, where
$S_{ij}$ is given by Eq.~(\ref{26}) with the ``full" $V(\omega,q)$.

Neglecting the disorder-induced damping of the dynamically screened
interaction yields
\be
S_{\rm ff}=S_{\rm fb}=t_c/2\tau_{ee}
\label{27}
\ee
and the exact cancellation of the self-energy and vertex parts in the total
dephasing action, hence the vanishing of $\tau_\phi^{-1}$ (\ref{24}) in the
clean limit. It is thus only because of the small difference between $S_{\rm
ff}$ and $S_{\rm fb}$ produced by the dressing of $V_{\mu\nu}(\omega,q)$ by
impurities (``dirty RPA" \cite{gornyi05,gornyi07}) that the dephasing action
(\ref{22}) is not zero. The characteristic energy transfer $\omega$ in the
processes that lead to the dephasing (i.e., contribute to the difference
$S_{\rm ff}-S_{\rm fb}$) is much larger than $\tau^{-1}$ [more accurately,
$\omega$ is spread over the range between $\tau_\phi^{-1}$ and $\tau^{-1}$,
because of the logarithmic factor in Eq.~(\ref{23})], which justifies the
expansion of $S$ in powers of $\tau^{-1}$, while the condition $T\tau_\phi\gg
1$ justifies the quasiclassical treatment of the electromagnetic fluctuations
in Eq.~(\ref{26}). Substituting Eq.~(\ref{24}) into Eq.~(\ref{14}) gives the
temperature scale $T_1\sim 1/\alpha^2\tau$ below which the localization
effects become strong (for the behavior of the conductivity at $T\ll T_1$ see
Ref.~\cite{fleishman80}). Note that $T_1\tau\gg 1$ for weak interaction.

\section{V. Energy relaxation}
\label{erel}

We now turn to the nonequilibrium properties of the disordered LL
\cite{bagopo}. Here we are primarily interested in the equilibration rate at
which an excited state relaxes to equilibrium (other aspects of the
nonequilibrium relaxation will be discussed elsewhere \cite{bagopo}). As
mentioned in Sec.~I, the integrability of the clean LL model precludes energy
relaxation. The absence of inelastic scattering in the LL deserves additional
comment. For scattering of electrons from different chiral branches on each
other, the energy and momentum conservation laws for linear electronic
dispersion lead to two equalities: $\omega-v_Fq=0$ and $\omega+v_Fq=0$. These
combine to give $\omega,q=0$ and thus no energy exchange [cf.\
Eq.~(\ref{20})]. For scattering of electrons of the same chirality $\mu$, the
energy-momentum conservation laws give a single equation $\omega-\mu v_Fq=0$
and at first glance the energy relaxation is allowed. Moreover, the relaxation
might seem to be very strong since the Golden-rule expression for the
probability of scattering contains the delta function $\delta (\omega-\mu
v_F)$ squared. For the point-like interaction, the diverging Hartree and
exchange terms cancel each other; however, for a finite-range
interaction---despite the LL model being still completely integrable---the
cancellation is no longer exact. The energy relaxation, nonetheless, is absent
in the LL model for a generic shape of the interaction potential. The point is
that beyond the Golden rule the diverging terms sum up to produce the
dynamically screened interaction between electrons (exactly given by the RPA),
which propagates with velocity $u(q)\neq v_F$. As a result, the probability of
scattering contains a product of two delta functions $\delta (\omega-\mu
v_Fq)\delta [\omega-\mu u(q)q]$ with {\it different} velocities, which yields
$\omega,q=0$ for electrons from the same chiral branch as well.

The energy relaxation is thus only present if one goes beyond the clean LL
model. One possibility comes from three-particle scattering \cite{Lunde},
which occurs for a nonzero range of interaction provided that the electronic
dispersion is nonlinear. The three-particle collision rate is small in the
parameter $T/\epsilon_F\ll 1$, where $\epsilon_F$ is the Fermi energy. Another
possibility is to take into account impurity backscattering, which may lead to
a much stronger mechanism of energy relaxation.

It is important that the nonequilibrium state of the LL in general cannot be
described in terms of a single---either bosonic or fermionic---distribution
function. The simplest example to illustrate this point is that of the clean
LL in which the left and right movers, separately at equilibrium within
themselves, are characterized by the Fermi distribution functions
$f^\pm_\epsilon=f_F(\epsilon-\mu_\pm)$ with different chemical
potentials. Then the distribution functions $N^\pm(\omega)$ of the left and
right plasmon modes are constructed as the convolutions of the fermion
functions:
\be
N^\pm(\omega)=\frac{1}{\omega}\int d\epsilon\,
f^\pm_\epsilon (1-f^\pm_{\epsilon-\omega})=N_B(\omega)~,
\label{Local}
\ee
i.e., are given by the {\it equilibrium} Bose distribution, independent of
$\mu_\pm$. This observation shows that the purely bosonic description of the
clean LL at a finite bias voltage is not complete. Such a ``partial
nonequilibrium" setup, in which the bosons are still at equilibrium, has been
studied previously by employing the conventional bosonization (see, e.g.,
Ref.~\cite{grabert}). Furthermore, the nonequilibrium transport through a {\it
single} impurity between {\it equilibrium} leads shifted by the voltage
$\mu_+-\mu_-$ has been studied in Ref.~\cite{fendley95}. However, the standard
scheme of bosonization will break down if the nonequilibrium distributions of
the injected right- and left-moving electrons are not the Fermi distributions.

The challenge is thus to formulate a theoretical framework to describe a
genuinely nonequilibrium situation in which both the bosonic and fermionic
excitations are out of equilibrium. It is worth stressing that in the
inhomogeneous case the nonequilibrium distribution functions do not obey the
simple local relation (\ref{Local}), since the distribution functions of
plasmons and electrons evolve with different velocities ($u$ and $v_F$).
Notice also that the necessity of introducing both the bosonic and fermionic
distribution functions is not peculiar to 1d: for higher-dimensional systems
see Ref.~\cite{Catelani}. 

Our approach to nonequilibrium phenomena in the LL uses as a base the
formalism of the ``functional bosonization", developed in Ref.~\cite{gornyi07}
for the treatment of disordered LL at equilibrium. A conceptually similar
formalism has been applied earlier for higher-dimensional disordered
conductors in the study of the single-particle density of states out of
equilibrium in Ref.~\cite{Gutman1}. The nonequilibrium tunneling density of
states in the clean LL has been considered within the functional bosonization
approach in Ref.~\cite{Gutman2}. Here we formulate the theory of the
disordered LL out of equilibrium, which builds on the approaches of
Refs.~\cite{Khmelnitskii,Kamenev} and Ref.~\cite{Catelani}, in terms of the
effective nonequilibrium real-time action. To account for the e-e interaction,
we introduce a dynamical field $\phi(x,t)$ which decouples the four-fermion
term in the action by means of the conventional Hubbard-Stratonovich
transformation.

The central object of our theory is the quasiclassical Green's function ${\hat
g}(x,t_1,t_2)$ for electrons in the LL, taken at coinciding spatial points
\cite{Shelankov}:
\bea
\hat{g}(x, t_1, t_2)&=&\lim_{x'\to x}\Bigl[\,2iv_F g(x,x',t_1,t_2) 
\nonumber \\
&-&{\rm sign}(x-x')\delta(t_1-t_2)\,\Bigr]~. 
\eea
This function, which is a $4\times 4$ matrix in the Keldysh and channel
(right/left) space, 
satisfies the Eilenberger equation \cite{Eilenberger}:
\be
iv_F\partial_x\hat{g}+[\,i\partial_t\hat\tau_z-\hat{H}\,,\hat{g}\,]=0~,
\label{Eilen_Eq} 
\ee
where 
\be
\hat{H}=\hat\phi\hat\tau_z+\frac{1}{2}({U_b}
\hat\tau^++{U_b^*}\hat\tau^-)~. 
\label{Eilen_Eq_1}
\ee
Here and throughout this section below, $v_F$ means the renormalized velocity
$v_F^*$ [see the discussion around Eq.~(\ref{5})], so that the difference
between $u$ and $v_F$ is of order $\alpha^2$ for small $\alpha$. 
Equation (\ref{Eilen_Eq}) describes the motion of an electron in the random
potential characterized by the backscattering amplitude $U_b(x)$
[Eq.~(\ref{6})] in the presence of the dynamic field $\hat\phi(x,t)={\rm
diag}\,(\hat\phi^+,\hat\phi^-)$, where
$\hat\phi^\mu(x,t)=\phi^\mu_1(x,t)+\hat\sigma_x\phi^\mu_2(x,t)$ and
$\phi_1^{\mu}$ and $\phi_2^{\mu}$ are the classical and quantum components of
the Hubbard-Stratonovich field with chirality $\mu$, respectively. The Pauli
matrices $\tau_z$ and $\tau^\pm=\tau_x\pm i\tau_y$ act in the channel
space. We also introduce the Pauli matrices $\hat\sigma_{x,y,z}$ that act in
the Keldysh space. The Hamiltonian $\hat{H}$ (\ref{Eilen_Eq_1}) is defined on
the direct product of the time, Keldysh, and channel spaces. Accordingly, the
commutator $[\,,\,]$ in Eq.~(\ref{Eilen_Eq}) is understood with respect to all
three (``discretized" time, Keldysh, and channel) indices. The operator
$\partial_t$ acts as $\overrightarrow\partial_{t_1}$ from the left and as
$(-\overleftarrow\partial_{t_2})$ from the right. For the case of linear
electronic dispersion, assumed in the LL model, the Eilenberger equation
(\ref{Eilen_Eq}) is {\it exact} for any given realization of the
backscattering amplitude $U_b(x)$.

The next step is to average Eq.~(\ref{Eilen_Eq}) over disorder. At this point
we disregard the localization effects \cite{gornyi05,gornyi07}, which limits
the applicability of the subsequent derivation to sufficiently high
(effective) temperatures; specifically, for the length of the quantum wire
larger than the mean free path to $T\gg T_1\sim 1/{\alpha^2\tau}$ (see the end
of Sec.~IV). Under this condition we can perform the disorder averaging at the
level of the self-consistent Born approximation, which gives
\bea
i\mu v_F\partial_x\overline{\hat{g}^\mu}+\Bigl[i\partial_t-
\hat\phi +\frac{i}{4\tau}\overline{\hat{g}^{-\mu}},
\overline{\hat{g}^\mu}\Bigr]=0
\label{Eilen_av}
\eea
for the disorder-averaged Green's function $\overline{{\hat g}^\mu}$. In
what follows we only deal with the averaged propagators and therefore omit the
bar for brevity.

The Green's function $\hat g$ satisfies the normalization condition
\be
{\hat g}\circ {\hat g}={\hat 1}\,\delta(t_1-t_2)~,
\label{Constrain}
\ee
where the dot denotes the convolution in all three (time, Keldysh, and
channel) spaces. The constraint (\ref{Constrain}) enables us to formulate the
effective action that reproduces Eq.~(\ref{Eilen_av}) as its saddle point in
the form essentially combining the actions derived in Refs.~\cite{Khmelnitskii}
and \cite{Kamenev}:
\bea
S[\,\hat{g},\hat\phi,{\hat A}\,]&=&-\frac{1}{2v_F}{\rm
Tr}\left[\,(i\partial_t-
\hat\phi)\hat\tau_z+v_F\hat{A}\,\right]\hat{g} \nonumber\\
&-&\frac{i}{2}{\rm Tr}\,\hat{g}_0T^{-1}\partial_xT-\frac{i}{8v_F\tau}{\rm
Tr}\,\hat{g}^+\hat{g}^-~. \nonumber\\
\label{ActionK}
\eea
The Green's function in Eq.~(\ref{ActionK}) is represented as
$\hat{g}=T\hat{g}_0T^{-1}={\rm diag}\,(\hat{g}^+,- \hat{g}^-)$, where
$\hat{g}_0= {\rm diag}\,(\hat{g}^+_0,-\hat{g}^-_0)$ corresponds to the saddle
point of the action of the noninteracting problem and the unitary
transformation $T$ (diagonal in the channel space) parametrizes possible
fluctuations around $g_0$ [satisfying the constraint (\ref{Constrain})],
induced by fluctuations of the field $\hat\phi(x,t)$. To generate the response
functions in the Keldysh formalism \cite{rammer86,kamenev04}, we have also
added the external-source term ${\hat
A}(x,t)=a_1(x,t)+\hat\sigma_xa_2(x,t)$. The trace operation includes the
summation over the Keldysh and space indices and the integration over
time. The Keldysh partition function of the system can now be expressed as a
functional integral over $\hat\phi$,
\bea
{\cal Z}[A]&\sim&\int {\cal D}\phi^{\mu}_{1,2}(x,t)\,\exp\left\{ iS[\,\hat\phi,
{\hat g},{\hat A}\,] \right. \nonumber \\ 
&+& \left. \frac{i}{2} {\rm Tr}\,\hat\phi\left(
V_f^{-1}\hat\tau_x+\frac{1}{2\pi v_F}\right)\hat\sigma_x\hat\phi\right\}~,
\eea
where ${\hat g}(x,t_1,t_2;[\hat\phi(x,t)])$ is understood as minimizing the
action ({\ref{ActionK}}) for a given $\hat\phi (x,t)$ under the constraint
(\ref{Constrain}).

Having written the Eilenberger equation (\ref{Eilen_av}) and its action
(\ref{ActionK}) we now use the standard technique \cite{rammer86,kamenev04} to
derive the quantum kinetic equations. We proceed at one-loop order with
respect to the effective interaction, which is equivalent to the ``dirty RPA"
\cite{gornyi07}. The one-loop derivation is controlled by the small parameters
$1/T\tau_\phi \ll 1$ and $\alpha \ll 1$. More specifically, following the
framework of Ref.~\cite{Catelani}, we introduce three different distribution
functions for each $\mu$. The first one, $f^\mu(\epsilon,x,t)$, describes the
{\it bare} electrons, moving with velocity $v_F$. The other two,
$N_p^\mu(\omega,x,t)$ and $N_g^\mu(\omega,x,t)$, describe two types of bosons,
having velocities $u_p=v_F/K$ and $u_g=v_F$. The bosons of the first kind
represent the usual plasmons ($p$) of the LL, whereas those of the second kind
are ``ghosts'' ($g$) constructed from the bare electron-hole pairs, thus
preventing from a double-counting of the degrees of freedom in the system [see
the discussion around Eqs.~(\ref{Energy_e})-(\ref{joule}) below].

We first apply the gauge transformation
\be
{\tilde g}^\mu (x,t_1,t_2)=e^{-i\hat\theta^\mu(x,t_1)}\hat{g}^\mu (x,t_1,t_2) 
e^{i\hat\theta^\mu (x,t_2)}~,
\label{gauge}
\ee
where $\hat\theta^\mu =\theta_1^\mu +\hat\sigma_x\theta_2^\mu$ has the same
Keldysh structure as the field $\hat\phi$. This transformation is similar to
that in Ref.~\cite{lerner04,eckern07}, but different in that the equation of
motion for the phase $\hat\theta$ in the field $\hat\phi$ will incorporate
disorder, see Eq.~(\ref{Linear_Rel}) below. The ``rotated" Green's functions
${\tilde g}^\mu$ are expressed in terms of the electron distributions
$f^\mu_\epsilon(x,t)$, written in the time domain, as
\be
{\tilde g}^\mu =\left[
\begin{array}{cc}
\delta(t_1-t_2) & 2h^\mu (t_1,t_2,x) \\
 0 & -\delta(t_1-t_2) 
\end{array} \right]~,
\ee
where $h^\mu =\delta(t_1-t_2)-2f^\mu (t_1,t_2,x)$,
\be
f^\mu(t_1,t_2,x)=\int\!{d\epsilon\over 2\pi}\,e^{i\epsilon 
(t_1-t_2)}f^\mu_\epsilon[x,(t_1+t_2)/2]~,
\ee
and we impose the condition 
\be
f^\mu (t_1,t_2,x)\vert_{t_1\to t_2}=\frac{i}{2\pi(t_1-t_2+i0)}~.
\label{cond}
\ee
The fast charge and current fluctuations are now encoded in the fluctuations
of the phase factors $e^{\pm i\hat\theta}$ in Eq.~(\ref{gauge}). The
gauge-transformed action reads
\bea
S[\hat\theta,\hat\phi ,{\tilde g}]&=&S_e+S_b+S_{\rm int}+S_{\rm imp}~,  
\label{ActionG} \\
S_e&=&-\frac{1}{2 v_F}{\rm Tr}\left(i\partial_t-{\hat L}_0\hat\theta 
-\hat\phi\right)\hat\tau_z{\tilde g} \nonumber \\
&&-\frac{i}{2}{\rm Tr}\,\hat{g}_0T^{-1}\partial_xT~, \\
S_b&=&\frac{1}{2\pi v_F}{\rm Tr}\left[\,\frac{1}{2}(\partial_t\hat\theta) 
\,{\hat L}_0\,\hat\sigma_x\hat\theta
+\hat\phi\,\hat\sigma_x\partial_t\,\hat\theta\right]~, \nonumber\\ \\
S_{\rm int}&=&\frac{1}{2}{\rm Tr}\,\hat\phi 
\left(V_f^{-1}\hat\tau_x+\frac{1}{2\pi v_F}\right)
\hat\sigma_x\hat\phi~, \\
S_{\rm imp}&=&-\frac{i}{8v_F\tau}{\rm Tr}\, 
e^{-i(\hat\theta^--\hat\theta^+)}{\tilde g}^+ 
e^{i(\hat\theta^--\hat\theta^+)}{\tilde g}^-~, \nonumber \\
\label{ActionG_imp} 
\eea
where ${\hat L}_0=\partial_t+\hat\tau_zv_F\partial_x$.

We treat the fluctuations of $\hat\theta$ and $\hat\phi$ in the Gaussian
approximation by expanding Eq.~(\ref{ActionG}) around the saddle point of $S$
for a given $\hat{g}^\mu$. Optimizing then the action with respect to
$\hat\theta$ for a given $\hat\phi$, we get a linear relation between
$\hat\theta$ and $\hat\phi$:
\be
{\hat D}^{-1}_g\theta=-\hat\sigma_x\phi~,
\label{Linear_Rel}
\ee
where we introduce the vector notation $\theta
=(\theta_1^+,\theta_1^-,\theta_2^+,\theta_2^-)^T$,
$\phi=(\phi_1^+,\phi_1^-,\phi_2^+,\phi_2^-)^T$, $T$ stands for
transposition, and the particle-hole propagator $D_g$ is constructed as
\be
{\hat D}^{-1}_g=(\partial_t+\hat\tau_zv_F\partial_x)\hat\sigma_x 
+\frac{1}{2}\hat\gamma (1-\tau_x)
\label{D_g} 
\ee
with
\bea
\hat\gamma&=&\frac{1}{\tau}\left(
\begin{array}{cc}
0 & -1\\
1 & 2B_\omega
\end{array} 
\right)~,
\label{Gamma} \\
B_{\omega}&=&\frac{1}{2\omega}\sum_{\mu}\int d\epsilon
\left(1-h^{\mu}_{\epsilon}\,h^{-\mu}_{\epsilon-\omega}\right)~.
\label{B_def}
\eea
The scattering operator $\hat\gamma$ (\ref{Gamma}) describes the
decay/recombination of the collective electron-hole excitation into/from the
electron and hole moving in opposite directions, assisted by impurity
scattering. Note that the approximation (\ref{Linear_Rel}) is equivalent
\cite{footnote} to the ``dirty RPA" \cite{gornyi07} in Sec.~IV.

Substituting Eq.~(\ref{Linear_Rel}) back into the approximate quadratic
action, we obtain the ``dirty-RPA" propagator of the effective interaction
\be
\langle\phi\phi^T\rangle=\frac{i}{2}\hat V 
=\frac{i}{2}\left(\hat\sigma_x\hat\tau_xV_f^{-1}
-\hat\Pi\right)^{-1}~, 
\label{Phi_Cor_1}
\ee 
where
\be
\hat\Pi=\frac{1}{2\pi v_F}\left[\,\hat\sigma_x
\left(\partial_t{\hat D}_g\right)\hat\sigma_x-\hat\sigma_x\,\right]
\label{Pi_operator}
\ee
is the polarization operator. By combining Eqs.~(\ref{Phi_Cor_1}) and
(\ref{Linear_Rel}) we get the correlator of the phases $\theta$ 
(cf.\ Ref.~\cite{levitov01})
\be
\langle\theta\theta^T\rangle=\frac{i}{2}\,
{\hat D}_g\,\sigma_x\,{\hat V}\,\sigma_x\,{\hat D}_g 
=-\frac{i\pi v_F}{\partial_t}\left({\hat D}_p-{\hat D}_g\right)~,
\label{Theta_Cor}
\ee
where ${\hat D}_p$ is the renormalized particle-hole propagator corresponding
to the plasmon modes with velocity $u$ given by Eq.~(\ref{5a}):
\be
{\hat D}^{-1}_p=\left({\partial_t\over
1+\alpha\hat\tau_x}+v_F\hat\tau_z\partial_x\right)\hat\sigma_x+{1\over
2}\hat\gamma (1-\hat\tau_x)~.
\ee

As follows from Eq.~(\ref{ActionG_imp}), the only phase that is coupled to the
electron backscattering off disorder is $\Phi=\frac{1}{2}(\theta^--
\theta^+)$. Another observation is that the propagators of the fluctuations of
$\theta$ have two different types of poles: $\omega=\pm uq$ and $\omega=\pm
v_F q$, both smeared by disorder. It is thus convenient to define the
correlator of the phase $\Phi$ as a difference of two terms:
\be
\langle\Phi\Phi^T\rangle =\frac{i}{2}\left(\hat{\cal L}_p  
-\hat{\cal L}_g\right)~,
\label{Phi_Cor}
\ee
where
\be
\hat{\cal L}_b=-\frac{i\pi v_F}{2\partial_t}\sum_{\mu\nu}\mu\nu
{\hat D}_b^{\mu\nu}
\label{Phi_Cor_2}
\ee
and $b=p,g$ denotes the plasmon and ghost modes, which differ from each other
in that the plasmon mode is characterized by velocity $u$, whereas the ghost
mode by velocity $v_F$. Then the Wigner-transform of the Keldysh correlator
$\langle\Phi\Phi^T\rangle_K$ has to be described by four different
distribution function, $N_p^\pm(\omega,x,t)$ and $N_g^\pm(\omega,x,t)$,
evolving with velocities $u$ and $v_F$ to the right and to the left:
\bea
&&\langle\Phi\Phi^T \rangle_K(\omega,q\simeq\pm\omega/u,x,t)= 
\label{Np}\\
&&\qquad\qquad\left[2N_p^\pm(\omega,x,t)+1\right]
{\rm Im}{\cal L}_p^A(\omega,q)~, \nonumber \\
&&\langle\Phi\Phi^T\rangle_K(\omega,q\simeq\pm\omega/v_F,x,t)= 
\label{Ng}\\
&&\qquad\qquad-\left[2N_g^\pm(\omega,x,t)+1\right]
{\rm Im}{\cal L}_g^A(\omega,q)~. \nonumber
\eea

To derive the kinetic equation for the electron distribution function, the
next step is to write down the equation of motion for the gauge-transformed
Green's function ${\tilde g}^\mu$ (\ref{gauge}). The latter follows from 
the relation
\be
\frac{\delta}{\delta{\tilde g}^\mu}\left(S_e+S_{\rm imp}\right)=0~.
\ee
Using Eq.~(\ref{Linear_Rel}) we represent $S_e$ as
\be
S_e=-\frac{1}{2v_F}{\rm Tr}\left(i\partial_t\hat\tau_z  
-\tilde\phi\right){\tilde g}~,
\ee
where the shifted phase $\tilde\phi =\tilde\phi_1+\tilde\phi_2\,\hat\sigma_x$,
\be
\tilde\phi_\alpha
=\sum_\beta (\hat\sigma_x\hat\gamma)_{\alpha\beta}\,\Phi_\beta~,
\ee
and $\hat\gamma$ is given by Eq.~(\ref{Gamma}). Notice that the field
$\tilde\phi$ does not depend on the chiral index $\mu$. The Eilenberger
equation for ${\tilde g}^\mu$ thus reads
\be
i\mu v_F\partial_x{\tilde g}^\mu+\left[i\partial_t-\mu\tilde\phi+ 
\frac{i}{4\tau}e^{2i\mu\hat\Phi} 
{\tilde g}^{-\mu}e^{-2i\mu\hat\Phi},\,\,{\tilde g}^\mu\right]=0~.
\label{Eilen_G}
\ee
Equation (\ref{Eilen_G}) has to be averaged over the fluctuations of the phase
$\Phi$ with the correlator given by Eq.~(\ref{Phi_Cor}). Within the
``dirty-RPA" it is sufficient to represent ${\tilde g}^\mu$ as a sum ${\tilde
g}^\mu=\langle{\tilde g}^\mu\rangle+\delta{\tilde g}^\mu$, where
$\langle{\tilde g}^\mu\rangle$ is the mean value, and take into
account only the term in $\delta{\tilde g}^\mu$ that is linear in the
fluctuations of $\Phi$, keeping in mind that the quadratic-in-$\Phi$
fluctuations of ${\tilde g}^\mu$ are incorporated in the mean value. 

By linearizing Eq.~(\ref{Eilen_G}) around the average
$\langle{\tilde g}^\mu\rangle$, we obtain $\delta{\tilde
g}^\mu=-2(\delta f^\mu)\,\hat\sigma_+$, where the fluctuation of the
distribution function obeys
\be
\sum_{\mu}\left[{\hat D}^{-1}_{g, R}(\omega)\right]^{\nu\mu}\, 
\delta f^\mu(\epsilon_1,\epsilon_2)= 
\frac{i\nu}{\tau}\,\sum_{\alpha=1,2}
\lambda^\nu_\alpha (\epsilon_1,\epsilon_2)\,\Phi_\alpha(\omega)~.
\label{dff}
\ee
In Eq.~(\ref{dff}), $\delta f^\mu (\epsilon_1,\epsilon_2)$ is the Fourier
transform of $\delta f^\mu(t_1,t_2,x)$ and $\omega=\epsilon_1-\epsilon_2$. The
source terms $\lambda^\nu_\alpha$ are expressed through the averages 
$h^\mu_\epsilon$ as
\bea
\lambda^\nu_1(\epsilon_1,\epsilon_2)&=&\frac{\nu}{2}\sum_\mu\mu\,
\left(h^{\mu}_{\epsilon_2}-h^{\mu}_{\epsilon_1}\right)~, \\
\lambda^{\nu}_2(\epsilon_1,\epsilon_2)&=&1+B_{\omega} 
\left( h^{\nu}_{\epsilon_2}-h^{\nu}_{\epsilon_1} \right) \\
&-&\frac{1}{2} h^{\nu}_{\epsilon_1} 
h^{-\nu}_{\epsilon_2}-\frac{1}{2} h^{\nu}_{\epsilon_2} 
h^{-\nu}_{\epsilon_1}~, \nonumber
\eea
where $h^\mu_\epsilon=1-2f^\mu_\epsilon$.  

Notice that the general formalism of the ``nonequilibrium functional
bosonization" [Eq.~(\ref{Eilen_G})] allows, in principle, for a
nonperturbative treatment of both the elastic renormalization and the
inelastic scattering if the phases $\Phi$ are kept in the exponents (see, in
particular, the calculation of the tunneling density of states in the clean LL
out of equilibrium in Ref.~\cite{Gutman2}). For our purposes in this paper, it
is sufficient to expand the exponential factors to second order in $\Phi$.

The Eilenberger-type equation for the average $\langle{\tilde g}^\mu
\rangle$ can now be written in the form
\be
i\mu v_F\partial_x\langle{\tilde g}^\mu\rangle+ 
\Bigl[i\partial_t+\frac{i}{4\tau} 
\langle \tilde g^{-\mu}\rangle ,
\,\langle{\tilde g}^\mu\rangle\Bigr]= 
\hat{\rm St}^\mu_{e-e}+\hat{\rm St}^\mu_{e-b}~.
\label{Eilen_av_1}
\ee
The collision integrals in the right-hand side of Eq.~(\ref{Eilen_av_1}) come
from the averages of second order in the fluctuations of $\Phi$. There are two
types of the collision integrals. One, $\hat{\rm St}^{\mu}_{e-b}$, comes from
the second-order terms in the expansion of the phase factors $e^{\pm
2i\mu\hat\Phi}$ in Eq.~(\ref{Eilen_G}). The other, $\hat{\rm St}^{\mu}_{e-e}$,
stems from the contraction of the linear correction $\delta{\tilde g}^\mu$
with the fluctuations of $\hat\Phi$.
 
The kinetic equation for the electron distribution function is obtained by
taking the Keldysh part of Eq.~(\ref{Eilen_av_1}):
\bea
&&[\partial_t+\mu v_F(\partial_x+e{\cal E}\partial_\epsilon)]f^\mu_\epsilon=
-\frac{1}{2\tau}(f^{\mu}_\epsilon-f^{-\mu}_\epsilon) \nonumber \\
&&\qquad +{\rm St}_{e-e}+\sum_{b=p,g}\left(\mu\,{\rm St}^{\rm el}_{e-b} 
+{\rm St}^{\rm inel}_{e-b}\right)~. 
\label{Fermions_KE}
\eea
The electron-boson collision integral ${\rm St}_{e-b}^{\mu}$ describes 
emission and absorption of the bosons of type $b=p,g$ by the fermions. Its
inelastic part, which is symmetric with respect to the channel index $\mu$, 
reads
\bea
&&{\rm St}_{e-b}^{\rm inel}(\epsilon)=\pm\frac{1}{4}\sum_{\mu\nu} 
\int_{-\omega_0}^{\omega_0}\!d\omega\,\omega {\cal K}_{e-b}(\omega) 
\nonumber \\
&&\times\left[N^{\nu}_b(\omega)f^{-\mu}_{\epsilon-\omega}
(1-f^{\mu}_\epsilon)-[1+N^{\nu}_b(\omega)]f^\mu_\epsilon 
(1-f^{-\mu}_{\epsilon-\omega})\right]~. \nonumber\\
\label{Bosons_inel}
\eea
The elastic (antisymmetric in $\mu$) part is given by
\bea
{\rm St}_{e-b}^{\rm el}(\epsilon)&=&\pm\frac{1}{4}\sum_{\mu\nu} 
\int_{-\omega_0}^{\omega_0}\!d\omega\,\omega {\cal K}_{e-b}(\omega) 
\nonumber \\
&\times&\mu\left[N^{\nu}_b(\omega)( f^{\mu}_{\epsilon}
-f^{\mu}_{\epsilon-\omega})+f^\mu_\epsilon 
f^{-\mu}_{\epsilon-\omega})\right]~,\nonumber\\ 
\label{Bosons_el}
\eea
where the signs $\pm$ refer to the plasmons (+) and ghosts $(-)$. In
Eqs.~(\ref{Fermions_KE})-(\ref{Bosons_el}), $\tau$ and $u$ are understood as
renormalized down to a scale $\omega_0$, so chosen that
$\Delta\epsilon\ll\omega_0\ll \Lambda$ but
$\alpha\ln\left(\omega_0/\Delta\epsilon\right)\ll 1$, where $\Delta\epsilon$
is a characteristic energy scale for the distribution function (for the
quantum wire biased by a voltage $V$ it is given by $\max\{T,eV\}$). The high
energy renormalization (see Sec.~III), coming from scales larger than
$\omega_0$ and independent of the details of the nonequilibrium state at low
energies, can thus be taken into account at the level of input parameters
($\tau$ and $u$) for the kinetic equations.

The e-e collision integral ${\rm St}_{e-e}$ describes inelastic fermionic
collisions due to the interaction via the plasmon waves:
\bea
{\rm St}_{e-e}(\epsilon)&=&\frac{1}{4}\sum_{\mu\nu}\int_{-\omega_0}^{\omega_0} 
d\omega d\epsilon'{\cal K}_{e-e}(\omega) 
\nonumber\\
&\times&\Bigl[
f^{-\nu}_{\epsilon'}(1-f^{\nu}_{\epsilon'-\omega})
f^{-\mu}_{\epsilon-\omega}(1-f^{\mu}_{\epsilon}) \nonumber \\
&-&f^{\nu}_{\epsilon'-\omega}(1-f^{-\nu}_{\epsilon'})
f^{\mu}_{\epsilon}(1-f^{-\mu}_{\epsilon-\omega})\Bigr]~.
\eea
Note that ${\rm St}_{e-e}$ does not depend on the chiral indices. The
collision kernels are written as
\bea
{\cal K}_{e-b}(\omega)&=&\pm\frac{v_F}{\omega^2\tau}\int\frac{dq}{2\pi}
{\rm Re}\sum_{\mu\nu}\mu\nu{D}_b^{\mu\nu}(\omega,q)~, \\
{\cal K}_{e-e}(\omega)&=&{\cal K}(\omega)+
{\cal K}_{e-g}(\omega)-{\cal K}_{e-p}(\omega)~,
\label{Kernels}
\eea
where
\be
{\cal K}(\omega)=-\frac{1}{\pi\omega}\int\frac{dq}{2\pi} 
\sum_{\mu\nu}\,
{\rm Re}\,D_g^{\mu\nu}(\omega,q)\,{\rm Im}\,V^{\nu\mu}_R(\omega,q)
\label{KernK}
\ee
and the explicit
form of the propagators $D^{\mu\nu}_g$ and $V^{\mu\nu}$ can be found in
Ref.~\cite{gornyi07} [see Eqs.~(A16),(A17),(A21)-(A23) there; note that
$D^{\mu\nu}_g$ corresponds to $v_FD_{\mu\nu}$ in Eqs.~(A16),(A17)].

\begin{figure}[t]
\includegraphics[width=3.0in]{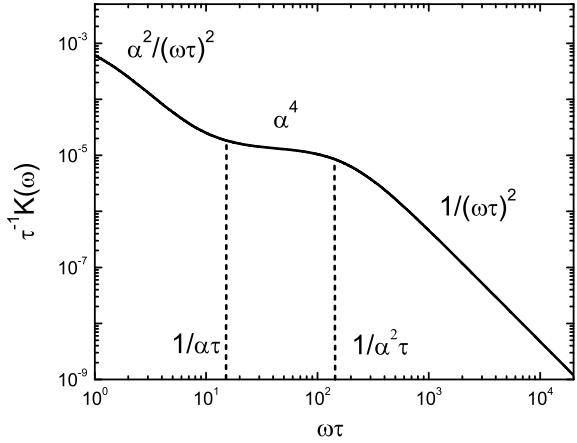}
\caption{
Frequency dependence of the collision kernel ${\cal K}(\omega)$
[Eq.~(\ref{KernK})] for $\alpha=0.05$. Three different types of
scaling behavior of the dimensionless product $\tau^{-1}{\cal K}(\omega)$ are 
indicated, as well as the characteristic values of $\omega$ .
} 
\label{Kw_plot}
\end{figure}

For the electron-boson terms we obtain in the ballistic limit of energy
transfers $\omega$ larger than $\tau^{-1}$ the simple expressions
\be
{\cal K}_{e-p}(\omega)=\frac{2K}{\omega^2\tau}~,\quad
{\cal K}_{e-g}(\omega)=\frac{2}{\omega^2\tau},\quad \omega\gg \tau^{-1}~.
\ee
The asymptotic behavior of ${\cal K}(\omega)$ in three parametrically
different ranges of $\omega$ in the limit of weak interaction $\alpha\ll 1$ is
as follows:
\bea
{\cal K}(\omega) = \left\{
\begin{array}{cc}
2\alpha^2/\omega^2\tau, &\tau^{-1}\ll\omega\ll\alpha T_1~, \\
8\alpha^4\tau, &\alpha T_1\ll\omega\ll T_1~, \\
2/\omega^2\tau, &\omega\gg T_1~, 
\end{array}\right.
\label{Komega}
\eea
where $T_1\sim 1/\alpha^2\tau$ is the characteristic temperature below which
the localization effects are strong (see the discussion at the end of
Sec.~IV). The log-log plot of ${\cal K}(\omega)$ in the whole range of
$\omega$ for a particular value of $\alpha$ (taken very small for the purpose
of illustration) is shown in Fig.~\ref{Kw_plot}.

An important feature of ${\cal K}(\omega)$ in Eq.~(\ref{Komega}) is its
nonperturbative behavior with respect to $\alpha$ at $\omega \gg \alpha
T_1\sim 1/\alpha\tau$. In particular, for $\omega\gg T_1$ the e-e collision
kernel does not at all depend on the e-e interaction strength. The origin of
the nonperturbative dependence on $\alpha$, despite $\alpha\ll 1$ being small,
is related to the analytical structure of ${\cal K}(\omega)$ in
Eq.~(\ref{KernK}). Specifically, the integrand of ${\cal K}(\omega)$ contains
eight poles, which in the limit of large $\omega$ are only slightly ``damped"
by disorder: $q \simeq \pm \omega(1\pm i/2\omega\tau)/v_F$ and $q \simeq \pm
\omega (1\pm i/2\omega\tau)/u$. As a result, the contour of integration in the
plane of $q$ is squeezed between two close poles ($u\to v_F$ for $\alpha\to
0$), one of which is in the upper-half plane and the other in the lower
one. At $\omega \gg T_1$ we thus have ${\cal
K}(\omega)\propto\alpha^2/|u-v_F|$ and $\alpha$ drops out altogether.

The kinetic equations for the bosonic distribution functions $N_{p,g}^\mu$ 
follow from Eqs.~(\ref{Phi_Cor}),(\ref{Np}),(\ref{Ng}):
\be
\left(\partial_t+\mu\,u_b\,\partial_x\right)N^\mu_b(\omega)=  
-\frac{1}{\tau}N^\mu_b(\omega)+{\rm St}_{b-e}(\omega)~,
\label{kineqbos} 
\ee
where 
\be
{\rm St}_{b-e}(\omega)=\frac{1}{2\omega\tau}\sum_{\mu}\int\ d\epsilon
f^{\mu}_{\epsilon}(1-f^{-\mu}_{\epsilon-\omega})
\label {Bosons_KE}
\ee
describes the creation of the boson from an electron-hole pair, where the
electron and hole move in the opposite directions.

The role of the ghost modes can be further elucidated if one considers the
energy conservation law. The electronic and bosonic energy
densities $\rho^{\epsilon}_e$, $\rho^{\epsilon}_b$ and the energy current
densities $j^{\epsilon}_e$, $j^{\epsilon}_b$ are given by
\bea
\rho^{\epsilon}_e&=&\frac{1}{2\pi v_F} 
\int_{-\infty}^\infty d\epsilon\,\epsilon (f^+_\epsilon+f^-_\epsilon)~, 
\label{Energy_e} \\
j^{\epsilon}_e&=&\frac{1}{2\pi}\int_{-\infty}^\infty d\epsilon \, 
\epsilon (f^+_\epsilon-f^-_\epsilon)~, \\
\rho^\epsilon_b&=&\frac{1}{2\pi u_b}\int_0^\infty d\omega \, 
\omega [N_b^+(\omega)+N_b^-(\omega)]~,
\label{Energy_b} \\
j^\epsilon_b&=&\frac{1}{2\pi}\int_0^\infty d\omega \,
\omega [N_b^+(\omega)-N_b^-(\omega)]~.
\eea
The prefactors in Eqs.~(\ref{Energy_e}) and (\ref{Energy_b}) represent the
density of states for electrons and bosons, respectively. Then the kinetic
equations (\ref{Fermions_KE}) and (\ref{kineqbos}) assure the conservation law
\be
\partial_t(\rho^{\epsilon}_e+\rho^{\epsilon}_p-\rho^{\epsilon}_g)
+\partial_x(j^{\epsilon}_e+j^{\epsilon}_p-j^{\epsilon}_g)=j_e\,{\cal E}~,
\label{joule}
\ee
where in the right-hand side $\cal E$ is the applied electric field
and $j_e$ is the induced current of charge. The kinetic energy without any
interaction is determined by the energy of the bare electrons. In the presence
of Coulomb interaction, the plasmon energy is given by a sum of the e-e
interaction energy and the kinetic energy of the bare electron-hole pairs, the
latter being the ghost energy by construction.  The total energy is thus given
by the sum of the plasmon and electron systems with a subtraction of the
energy of the ghosts.

We now turn to the rate of energy relaxation $\tau_E^{-1}$ in the limit of
weak nonequilibrium by linearizing the kinetic equations (\ref{Fermions_KE})
and (\ref{Bosons_KE}). One sees that at large energy transfers the inelastic
e-e scattering dominates over electron-boson collisions: ${\cal K}(\omega)\gg
{\cal K}_{e-g}(\omega)-{\cal K}_{e-b}(\omega)$ for $\omega\gg
1/\alpha^{3/2}\tau$.  Assuming that the large $\omega$ give the main
contribution to the energy relaxation in the limit $T\gg T_1$, where the
localization effects can be neglected (see Sec.~IV), the equilibration rate at
which the fermionic system relaxes to a locally equilibrium Fermi distribution
is estimated as
\be
\frac{1}{\tau_E(T)}\sim {1\over T}\int_0^T\!d\omega\,\omega^2{\cal K}(\omega)
\sim T^2{\cal K}(T)\sim\frac{1}{\tau}~. 
\label{Tau_E}
\ee 
Notice that the characteristic $\omega$ in Eq.~(\ref{Tau_E}) is of order $T$,
which justifies the use of ${\cal K}(\omega)$ only. On the other hand, this
makes it impossible to describe the equilibration in terms of the much simpler
Fokker-Planck equation in the energy space. Remarkably, the equilibration rate
(\ref{Tau_E}) does not depend on the strength of interaction and is given by
the elastic scattering rate. The interaction constant $\alpha$ enters only
through the condition $T\gg T_1$. The equilibration rate turns out to be much
smaller than the (clean) e-e collision rate~(\ref{19}) and also much smaller 
than the dephasing rate~(\ref{24}):
\be
\tau_E^{-1}\ll\tau_\phi^{-1}\ll\tau_{ee}^{-1}. 
\ee
It is also worth emphasizing that the characteristic energy transfers are
parametrically different in these three types of relaxation processes.

\section{VI. Summary}
\label{conc}

In this paper, we have studied the relaxation properties of interacting
spinless electrons in a disordered quantum wire within the Luttinger-liquid
model. We first review the basic concepts in the disordered Luttinger liquid
at equilibrium, including the elastic renormalization, dephasing, and
interference-induced localization. We have introduced the general framework
for describing the relaxation processes in the strongly correlated (non-Fermi)
electron system at nonequilibrium. Our main result is the coupled set of the
kinetic equations for the fermionic (\ref{Fermions_KE}) and bosonic
(\ref{kineqbos}) distribution functions. The peculiarity of the Luttinger
liquid model is that the electron-electron scattering rate (the inverse
lifetime of the fermionic excitations) is finite at nonzero temperature but
the energy exchange is exactly zero in the clean limit. The energy relaxation
occurs only due to the scattering off disorder. We have calculated the energy
equilibration rate that turns out to be independent of the strength of
electron-electron interaction at sufficiently high temperature, when the
Anderson localization effects are suppressed, and equal to the rate of elastic
scattering off disorder.

\section{Acknowledgments}
\label{acknow}

We thank D.~Gutman, A.~Kamenev, M.~Kiselev, Y. Nazarov, and A.~Yashenkin for interesting
discussions. The work was supported by the Center for Functional
Nanostructures of the Deutsche Forschungsgemeinschaft, by the Russian
Foundation for Basic Research, and by the Program ``Leading Russian Scientific
Schools". The work of DAB and IVG, conducted as part of the project ``Quantum
Transport in Nanostructures" made under the EUROHORCS/ESF EURYI Awards scheme,
was supported by funds from the Participating Organisations of EURYI and the
EC Sixth Framework Programme.


\begin{thebibliography}{1}
\bibitem{footnote} In Refs.~\cite{gornyi05,gornyi07} the propagator
$\hat{D}_g$ in Eqs.~(\ref{Linear_Rel}) and (\ref{Theta_Cor}) does not contain
disorder [Eq.~(\ref{20a})] and reduces to $L_0^{-1}$, which is, however,
sufficient for the calculation of $\tau_\phi$.
\end{thebibliography}
\end{document}